\documentclass{appolb}
\usepackage{graphicx}
\usepackage{dcolumn}% Align table columns on decimal point
\usepackage{bm}% bold math
\usepackage{hyperref}% add hypertext capabilities
\usepackage{xcolor}
 \usepackage{amsmath}
\begin{document}
\title{Fluctuations of conserved charges in strong magnetic fields from lattice QCD%
\thanks{Presented at the XXIXth International Conference on Ultra-relativistic Nucleus-Nucleus Collisions.}%
}
\author{Heng-Tong Ding, Sheng-Tai Li,
Jun-Hong Liu$^\dag$,
Xiao-Dan Wang
\address{Key Laboratory of Quark and Lepton Physics (MOE) and Institute of Particle Physics,
Central China Normal University, Wuhan 430079, China}
}

\footnotetext{\textit{$^\dag$~speaker. }}
\maketitle
\begin{abstract}
We present the first lattice QCD results of the second order fluctuations of and correlations among net baryon number, electric charge and strangeness in (2+1)-flavor lattice QCD in the presence of a background magnetic field with physical pion mass $m_{\pi}=135$ MeV. To mimic the magnetic field strength produced in the early stage of heavy-ion collision experiments we use 6 different values of the magnetic field strength up to $ \sim $10$m_{\pi}^2$. 
We find that the correlations between baryon number and electric charge along the transition line are substantially affected by magnetic fields in the current $eB$ window, which could be useful for probing the existence of a magnetic field in heavy-ion collision experiments.
\end{abstract}
  
\section{Introduction}
QCD phase structure in the nonzero magnetic fields has recently gained a lot of attention as the strong magnetic field is expected to be present in the early stage of peripheral heavy-ion collisions \cite{KHARZEEV2008227, SKOKOV_2009, Deng_2012}, early universe \cite{VACHASPATI1991258} and magnetars \cite{Enqvist_1993}. The chiral magnetic effect might be observed in heavy-ion collisions experimental observations if the magnetic field lasts long enough \cite{Kharzeev_2016, Kharzeev_2020}. The lifetime of the magnetic field  highly depends on the medium electrical conductivity, which is difficult to determine due to the inverse problem in first-principles calculations \cite{Astrakhantsev_2020, Ding_2016, Ding_2011}.Many efforts have been made to search for the signal of a magnetic field in the heavy-ion collision experiments\cite{PhysRevLett.123.162301,PhysRevLett.125.022301,PhysRevLett.121.132301,PhysRevLett.121.212301,STAR:2021mii}. Fluctuations of and correlations among net baryon number (B), strangeness (S), and electric charge (Q) have been very useful to probe the changes of degrees of freedom 
in QCD as well as the QCD phase structure at zero magnetic fields\cite{Luo:2017faz,Ding:2015ona}, as they are both theoretically computable and experimentally measurable. On the other hand, the fluctuations and correlations of B, Q, and S in nonzero magnetic fields have rarely been studied. Most of the published studies are based on the hadron resonance gas model\cite{Fukushima:2016vix,Ferreira:2018pux,Bhattacharyya:2015pra,Kadam:2019rzo} and Polyakov-Nambu-Jona-Lasinio model\cite{Fu:2013ica}. In our previous study, we present the lattice result of the fluctuations and correlations of B, Q, and S in nonzero magnetic fields for the first time where the computation was done using the highly improved staggered fermions with pion mass about 220 MeV\cite{Ding:2021cwv}. We found that the fluctuations and correlations of B, Q, and S are strongly affected by the magnetic fields. In this paper, We extend the lattice QCD studies to the case with physical pion mass $m_{\pi}$=135 MeV, and focus on a smaller temperature interval around the pseudo-critical temperature ranging from 0.9 $T_{pc}$ to 1.1 $T_{pc}$. To mimic the magnetic field strength produced in the early stage of heavy-ion collision experiments we now have 6 different values of the magnetic field strength $eB$ up to 10$m_{\pi}^2$.

\section{SECOND ORDER FLUCTUATIONS OF
CONSERVED CHARGES AND COMPUTATIONAL SETUP}
To calculate the fluctuations of conserved charges and their correlations in a thermal medium, the starting point is the pressure $p$ expressed in terms of the logarithm of partition function $Z$ as follows
\begin{equation}
\frac{p}{T^{4}} \equiv \frac{1}{V T^{3}} \ln Z\left(V, T, \mu_{\mathrm{B}}, \mu_{\mathrm{S}}, \mu_{\mathrm{Q}}\right),
\end{equation}
where the baryon (B), strangeness (S) and electric charge (Q) chemical potentials have following relations with the quark chemical potentials $\mu_u$, $\mu_d$ and $\mu_s$,
\begin{equation}
\begin{aligned}
\mu_{u} &=\frac{1}{3} \mu_{\mathrm{B}}+\frac{2}{3} \mu_{\mathrm{Q}}, \\
\mu_{d} &=\frac{1}{3} \mu_{\mathrm{B}}-\frac{1}{3} \mu_{\mathrm{Q}}, \\
\mu_{s} &=\frac{1}{3} \mu_{\mathrm{B}}-\frac{1}{3} \mu_{\mathrm{Q}}-\mu_{\mathrm{S}}.
\end{aligned}
\end{equation}
The fluctuations of the conserved charges and their correlations can be obtained by taking the derivatives of pressure with respect to the chemical potentials from lattice calculations evaluated at zero chemical potentials
\begin{equation}
\begin{aligned}
\hat{\chi}_{i j k}^{u d s} &=\left.\frac{\partial^{i+j+k} p / T^{4}}{\partial\left(\mu_{u} / T\right)^{i} \partial\left(\mu_{d} / T\right)^{j} \partial\left(\mu_{s} / T\right)^{k}}\right|_{\mu_{u, d, s}=0}, \\
\hat{\chi}_{i j k}^{\mathrm{BQS}} &=\left.\frac{\partial^{i+j+k} p / T^{4}}{\partial\left(\mu_{\mathrm{B}} / T\right)^{i} \partial\left(\mu_{\mathrm{Q}} / T\right)^{j} \partial\left(\mu_{\mathrm{S}} / T\right)^{k}}\right|_{\mu_{\mathrm{B}, \mathrm{Q}, \mathrm{S}}=0} .
\end{aligned}
\end{equation}
Here in our study we focus on the computation of quadratic fluctuations and correlations, i.e. $i+j+k = 2$. The expressions of quadratic fluctuations $\hat{\chi}_{i j k}^{\mathrm{BQS}}$ in terms of $\hat{\chi}_{i j k}^{u d s}$ can be easily obtained via Eq. 2.

The magnetic field is introduced along the $z$ direction, and is described by a fixed factor $u_{\mu}(n)$ of the U(1) field which is introduced in the Landau gauge~\cite{Bali_2012,DElia:2021yvk}. In our current simulations of (2+1)-flavor QCD in nonzero magnetic fields, the highly improved staggered quarks are adopted~\cite{Ding:2020hxw}.
The strange quark mass and the light quark mass are fixed to their physical value $m_s^{phy}$ and $m_l^{phy}$. To perform simulations at nonzero temperature, we change the lattice spacing $a$ at fixed $N_{\tau}$ to have different temperatures. 
Values of $N_{\tau}$ are fixed to 8 and 12. The temperature ranges from $\sim 0.9 T_{p c}$ to $\sim 1.1 T_{p c}$. The values of magnetic flux $N_b$ are 1,2,3,4 and 6 which correspond to the magnetic field strength $eB<$10$m_{\pi}^2$. 
Since the lattice spacing $a$ is not a constant, the interpolations of lattice data at different $T$ and $N_b$ are needed to have constant magnetic field strength in physical units among different temperatures as $a$ varies with temperature.

\section{Results and discussions}
		\begin{figure*}[h!]			
		\includegraphics[width=0.485\textwidth]{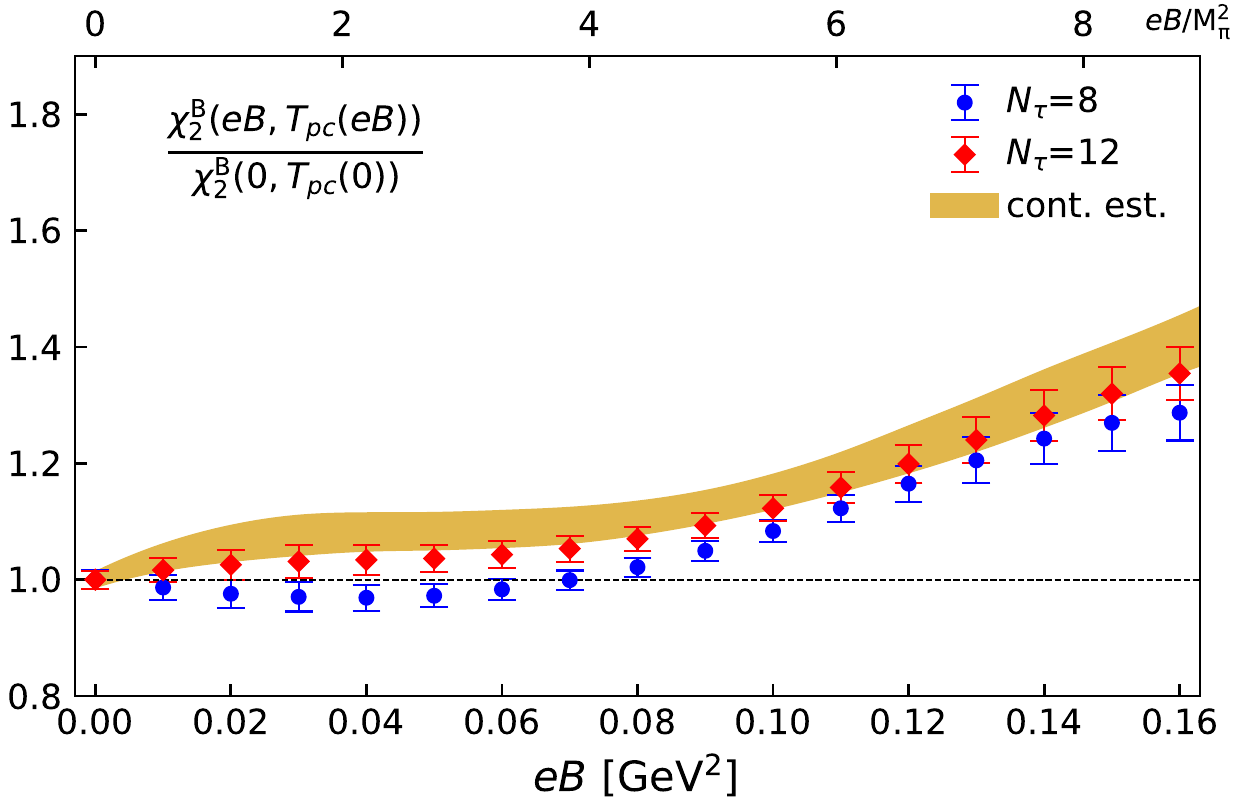}
		\includegraphics[width=0.485\textwidth]{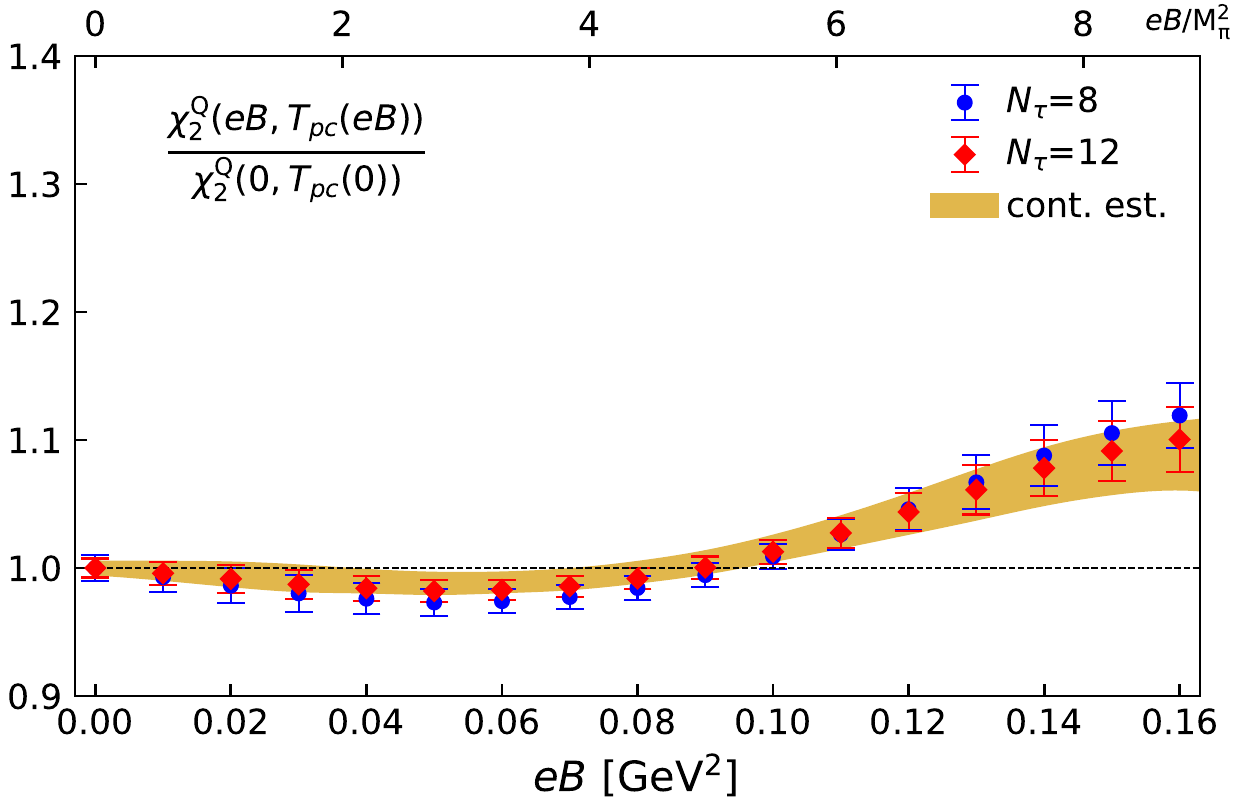}\\
		\includegraphics[width=0.485\textwidth]{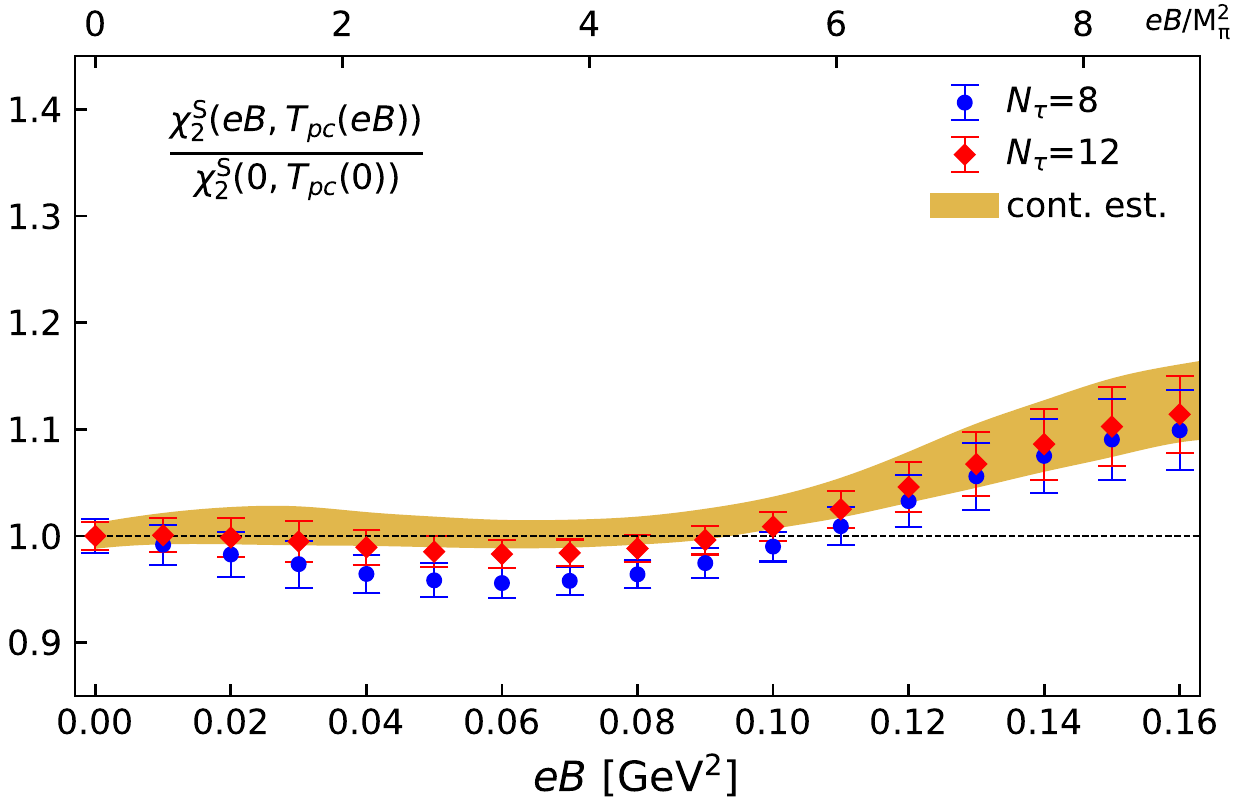}
		\includegraphics[width=0.485\textwidth]{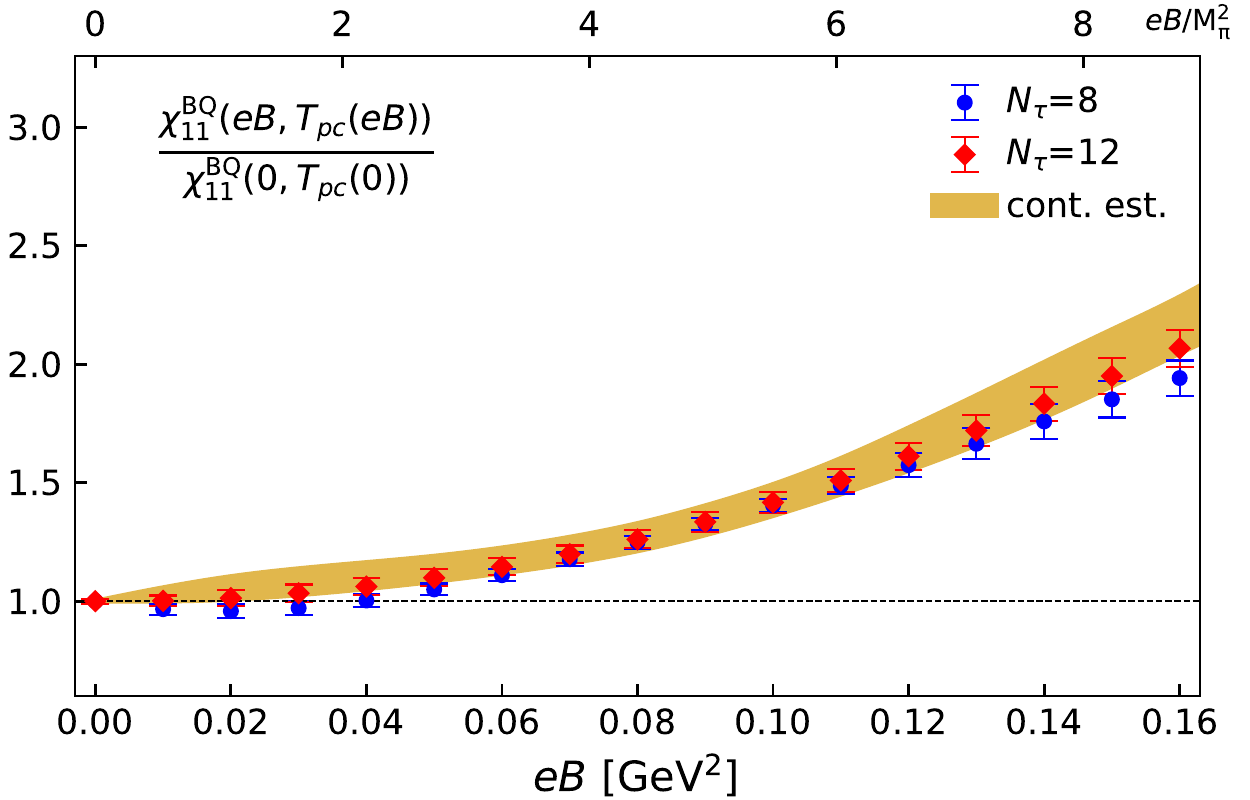}\\
		\includegraphics[width=0.485\textwidth]{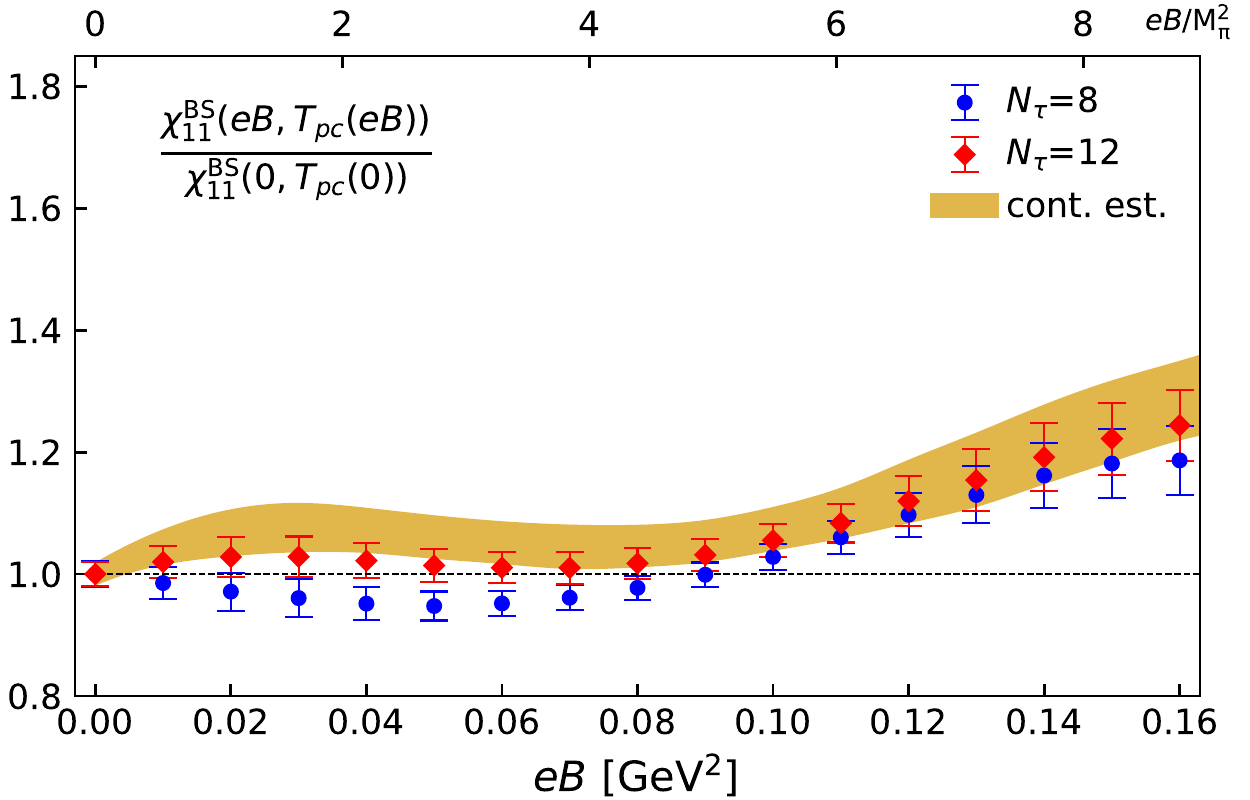}
		\includegraphics[width=0.485\textwidth]{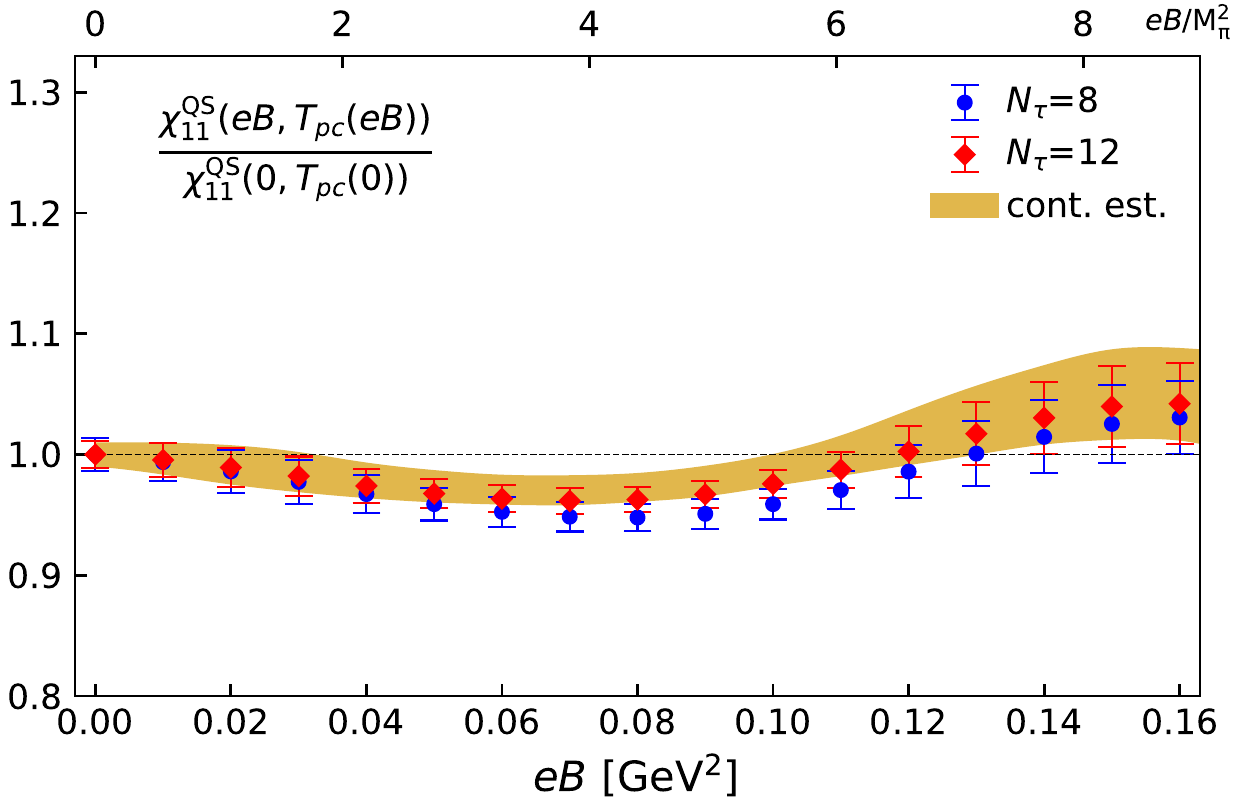}
		\caption{Normalized 2nd order fluctuations of and correlations among baryon number ($\mathrm{B}$), electric charge ($\mathrm{Q}$) and strangeness ($\mathrm{S}$) along the transition line $T\equiv T_{pc}(eB)$. The continuum estimates are denoted by the yellow band based on the lattice QCD data obtained on $N_{\tau}=8$ (blue points) and $N_{\tau}=12$ (red points) lattices.}
\label{fig:Q_along_tline}
	\end{figure*}

In heavy-ion collision experiments, the initial magnetic field strength varies from central collisions to peripheral collisions. Therefore, we would like to propose an $R_{cp}$-like quantity to detect the existence of magnetic fields in heavy-ion collisions. $R_{cp}(X)$ means the ratio of a certain quantity $X$ obtained in central collisions (weak/vanishing magnetic field) to that in peripheral collisions (strong magnetic fields). To achieve this we normalize the observable along the transition line by the
value of observable at vanishing magnetic fields: $X\left(eB, T_{pc}(eB)\right)/X\left(0,T_{pc}(0)\right)$.

In figure \ref{fig:Q_along_tline}, we show the $eB$ dependence of the normalized 2nd order fluctuation and correlations of baryon number, electric charge and strangeness along the transition line. These values equal to unity at $eB=0$ by definition, and the deviation from unity thus show the magnitude of the influence induced by the magnetic field. From figure \ref{fig:Q_along_tline} we can see that net baryon number related quantity: $\chi_2^{\rm B}$, $\chi_{11}^{\rm BS}$ and $\chi_{11}^{\rm BQ}$ are considerably influenced by the magnetic field. In particular, $R_{cp}$-like quantity for  $\chi^{\rm BQ}_{11}(eB,T_{pc}(eB))/ \chi^{\rm BQ}_{11}(0,T_{pc}(0))$ 
 reaches $\sim$2 at $eB\sim9 m_{\pi}^2$. This is a significant effect which might be useful for probing the existence
of a magnetic field in heavy-ion collision experiments. 
We remark here that our current results serves as a QCD baseline for the behavior of quadratic fluctuations
and correlation of conserved charges in external magnetic fields, and further studies are needed to connect to experimental observations.

\section*{Acknowledgement}
We thank Xiaofeng Luo, Swagato Mukherjee and Nu Xu for useful discussions. This work was supported by the NSFC under grant No. 11775096 and the Guangdong Major Project of Basic and Applied Basic Research No. 2020B0301030008. The numerical simulations have been performed on the GPU cluster in the Nuclear Science Computing Center at Central China Normal University ($\mathrm{NSC}^{3}$) and Wuhan Supercomputing Center. 
%uncomment the following lines to place a figure
%\begin{figure}[htb]
%\centerline{%
%\includegraphics[width=12.5cm]{Fig1}}
%\caption{Plot of ...}
%\label{Fig:F2H}
%\end{figure}
\bibliographystyle{unsrt}
\bibliography{QM2022_liu}

\end{document}